\documentclass[leqno,12pt]{article}

\usepackage[english]{babel}
\usepackage[utf8]{inputenc}
\usepackage{graphicx}
\usepackage[T1]{sansmath} 
\usepackage{array}
\usepackage{multirow}
\usepackage{booktabs}
\usepackage{colortbl}

\usepackage{hyperref}
\usepackage{longtable}
\usepackage{lscape}
\usepackage{amsmath}
\usepackage{afterpage}
\usepackage{setspace}
\usepackage[flushleft]{threeparttable}

\usepackage[a4paper,top=2.54cm,bottom=2.54cm,left=2.54cm,right=2.54cm]{geometry}

\usepackage{natbib}
\usepackage{breakcites}

\newcommand{\M}[1]{\mathbf{#1}}

\title{Analysis of Price and Income Elasticities of Energy Demand in Ecuador: A Dynamic OLS Approach}
\author{Kathia Pinzón \\
Escuela Politécnica Nacional\\
kathia.pinzon@epn.edu.ec}
\date{}
\providecommand{\keywords}[1]{\textbf{Keywords:} #1}
\providecommand{\jelcode}[1]{\textbf{JEL classification:} #1}

\begin{document}

\maketitle

\begin{abstract}
Energy consumption (EC) in Ecuador has increased significantly during the last decades, affecting negatively the financial position of the country since (1) large EC subsidies are provided in its internal market and (2) Ecuador is mostly a crude-oil exporter and oil-derivatives importer country. This research seeks to state the long-run price and income elasticities of energy demand in Ecuador, by analyzing information spanning the period 1970--2015. A co-integration analysis and an estimation by using a Dynamic Ordinary Least Squares (DOLS) approach considering structural breaks is carried out. Results obtained are robust and suggest that in the long-run energy demand in Ecuador (1) is highly income elastic, (2) has no relationship with its price and (3) has an almost unitary but inverse relationship with the industrial production level. Conclusions and economic policy suggestions are also provided. 
\end{abstract}

\keywords{Ecuador, energy demand, elasticity, co-integration, DOLS, structural break.}

\jelcode{Q43, Q48, O13.}

\section*{Introduction}

Energy consumption (EC) is implied in every economic activity: in activities related to production as well as in those related to consumption; so that it is one of the major factors involved in the economic system. Furthermore, it is demanded in such a recursive way that, not only the enterprises demand energy for carrying out their activities, but also those enterprises' performance allows people to demand more energy goods, encouraging in turn even greater production levels. \protect\cite{a}, to this respect, even mentions that with an expansion in the economy, the production increases over time, resulting in greater energy requirements to sustain the pace of development. Therefore, the EC in a given country is directly related too its economic performance. 

According to information published in BP\footnote{``BP Global'' is one of the world's leading integrated oil and gas companies. One of its key reports is the ``BP Statistical Review of World Energy'', which provides information regarding global energy trends and projections.} Statistical Review of World Energy 2016, the world EC has kept an increasing trend over the last 25 years, showing a mean annual growth rate of 1.9\%. In 2015 it stood at 13147.3 Mtoe.\footnote{Million tonnes oil equivalent. (Mega-toe). It is a unit of energy. A tonne of oil equivalent (toe) is the amount of energy released by burning one tonne of crude oil; though there are several kinds of crude oil which have different calorific values, the exact value is defined by convention so that 1 toe. equals approximately to 42 Gigajoules (GJ).} (1\% greater than in 2014) showing the same trend that the World GDP (At Purchasing Power Parity, measured in U.S. dollars at constant prices of 2011) presented: this latter grew at a mean rate of 3.37\% during the same period according to the World Bank. Most of countries, in fact, present growing trends in EC; however, some of them---mostly in Europe and Eurasia---keep their levels relatively constant. EC is, therefore, induced differently by countries worldwide. Moreover, as \protect\cite{b} mentions, such differences depend on factors like GDP level, industrial structure, lifestyle of citizens, geographical location and energy prices (especially relative energy prices).  

In several countries around the world, the growth in EC has led to economic growth \protect\citep[see, for instance][]{CO2Africa,ECMulticountry}. Nonetheless, it has also had negative ecological and financial effects in many countries worldwide. Firstly, its negative environmental impact has been expressed in an upward trend of CO2 emissions \protect\citep{CO2Malas,ECCO2,CO2Turkey,EUCO2China}. According to information of the BP Statistical Review of World Energy 2016, alike the EC, the world CO2 emissions level has also maintained an increasing trend (with a mean growth rate of 2.2\%) over the last 50 years. This has been mainly caused by the consumption of energy from fossil fuels, which are the most polluting among the existent ones\footnote{The amount of CO2 produced when a fuel is burned is a function of the carbon content of the fuel, so that the fossil fuels such as the coal, the oil and the natural gas are those which emits more CO2  per unit of energy output or heat content} and whose participation in the total world EC, according to the World Bank, was about the 80\% in 2013. Such adverse ecological effects, either directly or indirectly, generate costs to be covered by governments, not just in the present but also in the future (e.g., investment in public health, nature preservation, among others). Secondly, nowadays in several countries this is also causing budget disequilibrium difficulties due to the subsidies at which some of the kinds of energy are subject in the internal markets, in addition to the prices volatility of such goods at international level. In Latin America and the Caribbean, for instance, where energy use has also increased significantly over the last years (at a mean growth rate of about 3.9\% between 1965 and 2015), the costs generated due to negative externalities of energy use reached in the period 2011--2013 about the 2\% of GDP. Additionally, the costs generated by providing fuel and electricity subsidies reached about 1.8\% of GDP in the same period \protect\citep{ESiLA}.

In Ecuador, the scene is pretty similar. EC in this country is barely about the 2\% of the total in Latin America and the Caribbean EC. Nevertheless, while its GDP in constant terms has been increasing over the last 45 years at a mean rate of 1.76\%, EC has also been increasing at a mean rate of 6\%. This behavior is reasonable given that developing countries tend to present levels of growth in EC higher than their levels of economic growth \protect\citep{a}. However, such trend of EC, along with the strong dependence of the Ecuadorian economic system on oil-based fuels consumption and the existence of large indirect subsidies oriented to such fuels, have affected negatively the financial position of Ecuador since they have supposed an each time greater assignment of monetary resources---by the government---to fund EC: According to information from the Finances Ministry of Ecuador, while in 2011, the contribution of the Central Government to import oil-derivatives was of \$145,9 millions (6.2\% of the total DDFA\footnote{In 2008, given the great gap existent between the volumes of national demand and supply of oil derivatives in the country, an account called ``Deficit Derivatives Financing Account (DDFA)'' was created as part of the General State Budget, with the objective  of keeping permanently the necessary funds to import oil derivatives in  order  to cover the internal demand. Such account gathers (1) a transfer of the income perceived from oil derivatives sells done by the public enterprise ``PETROECUADOR'', (2) a transfer done by the Central Government and (3) a transfer  regarding revenues from certain exports of crude oil.}), in 2015 it was of \$1252.77 millions and represented the 27.1\% of the total DDFA.    

As \protect\cite{d} mentions, spurred by the oil price shocks in late 1973 and during the period 1979 to 1980, a lot of attention was devoted to the analysis of energy demand as a consequence of the dramatic events in energy markets and the increasing importance of this sector in national economies. Thus, it was performed a great effort to estimate the relationship between energy demand and factors such as income and energy price. Nevertheless, in Ecuador the research aimed to this respect has been lesser than in many other countries worldwide. At the  best of  my knowledge, in this country no work of this specific kind has been carried out before, in spite of the budget problems that the current upward trend of energy demand represents. 

Nowadays, the behavior of energy prices internationally is affecting again to Ecuador as well as to several other countries worldwide. In this country, the increasing energy demand along with the high volatility of oil prices in the external market, had generated high costs to the government given the existing trend to import some types of secondary energy and the significance of energy subsidies in its economy. This finally states the solution of the issues related to energy usage as one of the most important topics to consider about economic policy. In consequence, it becomes essential to know how sensitive energy demand is to changes in people's income and energy prices, in order to provide right responses of economic policy that conducts the country to a more ecologically and financially sustainable growth.   

This paper attempts to determine the price and income elasticities of energy demand in Ecuador by applying a Dynamic Ordinary Least Squares (DOLS) approach. The results will allow us to define important conclusions in order to shed lights on the adequate economic policy on this issue. Obtained results are robust. The reminder of this paper is as follows. Section \ref{tandecontext} states the theoretical and empirical context of the research. Section \ref{background} states the background of energy demand in Ecuador. Section \ref{methodology} explains the methodology applied and data used.  Section \ref{results} details the results obtained and their interpretation. Section \ref{conclusions} states the conclusions and policy implications of the research.

\section{Theoretical and Empirical Context} \label{tandecontext}

The demand of energy plays an important role in the economic system: changes in level or structure of EC could lead to significant changes in other macroeconomic variables in an economy. In analyzing energy usage and policy oriented to such issue, the whole energy demand specification represents a crucial input \protect\citep{d}. Price and income elasticity of energy demand, therefore, represent important tools in searching to generate important changes in the economic performance of a given country. Certainly, not only income and energy price are factors that define the amount of energy demanded, but also others such as industrial structure, resource endowments, etc \protect\citep{b,d}. However, most of the studies carried out to this respect had highlighted their importance especially because of their usefulness as policy economic tools at a macro level and, in some cases, due to the lack of data regarding other factors. 

Price and income elasticity of energy demand shows the level of responsiveness of the demanded amount of energy in respect to changes in its price and in income level, respectively. In the existent literature, energy demand is accepted to tend to be price inelastic due to its basic nature necessity and its lack of substitutes, whereas the income elasticity tends to vary across countries depending on the specific conditions of each one. Additionally to this respect, \protect\cite{b} point out that in developing countries energy demand tends to have a higher price elasticity than in developed ones, given the higher dependence of their developing industries on EC.  

Given the importance of energy demand not only in terms of clearly economic aspects but also in environmental ones, a considerable amount of research work has been carried out regarding this topic. However, at my knowledge, most of them have been developed regarding non-latinamerican countries. In fact, as \protect\cite{c} mentions, in the literature, most of the studies have been conducted by developed countries and mention vaguely the impact of oil prices on energy demand for developing countries. Besides, several of them are typically based on cross section analysis and therefore, offer only a representative measure or benchmark for price and income elasticity of energy demand in developing countries. 

Under such consideration, at the international level, for instance, one can quote several authors who had researched about this regard, in reference to developing countries as well as to developed ones \protect\citep[see, for example][]{a,b}. Several researches have looked for stating the estimates of energy demand elasticities for specific countries \protect\citep[see, for example][]{a,c}, meanwhile other ones have analyzed those estimators for groups of countries \protect\citep[see, for example][]{b,d,e} 

\protect\cite{a}, for instance, in his research, found the price and income elasticity of commercial energy demand in Nepal to be high (-1.65 and 3.04 using variables in levels, respectively); the petroleum energy demand to have a price elasticity between -0.83 and -0.55 and an income elasticity between 0.02 and 0.54; and the electric energy demand to have a price elasticity between -1.14 and -1.25 and an income elasticity of 4.51. This, by applying a distributed polynomial lag model with information spanning the period 1980--1999. It is worth mentioning that by the time at which such research was carried out (\protect\citeyear{a}), the Nepalese economy depended at a low level on commercial energy i.e.\ obtained from resources such as oil, coal, natural gas and the atoms' core (nuclear energy). Thus, the use of non-commercial energy sources (which are not easily measurable) allowed a higher sensitivity of commercial energy demand to prices.  

By other side, \protect\cite{c} determined that the demand for energy is fairly income elastic and price inelastic (50\% and -20\% on average, respectively) in the long run in Barbados, by applying a co-integration approach and analyzing the period from 1960 to 2005. 

\protect\cite{b}, besides, determined the price and income elasticities, both for the short and long run, for many countries of ASEAN and some East Asian countries by using a log-linear energy demand model. They found that the price elasticity is higher in developing countries than in developed ones and the income elasticity is relatively high (greater than 1) in most of those countries. Moreover, they concluded that price elasticity tends to be higher in countries where energy subsidies do not exist.

\protect\cite{d} found, by analyzing a group of 16 developing countries for the period from 1978 to 2003 and employing a dynamic heterogeneous panel estimation, that energy price and income elasticity for those countries in the long run are 0.03 and 0.17, respectively, while the short run ones are 0.02 and 0.1, respectively. They also showed that industrialization degree and CO2 emission levels are factors that affect significantly energy demand in those countries. 

Additionally, several investigations about this topic have been focused on specific kinds of energy. \protect\cite{e}, for instance, estimated the price and income elasticities of the residential electricity demand, by applying a panel co-integration approach, for a set of 18 OECD countries spanning the period 1978--2008. They concluded that demand in most of the countries is price inelastic, suggesting that changes in prices for seeking electrical energy conservation would fail as measures of economic policy. 

In Ecuador, few investigations about this topic have been carried out, and have been related only with specific kinds of EC. The research of \protect\cite{h}, for instance, estimates the price elasticity of electric energy residential demand in the ``Concession Area of the Center-South Regional Electrical Enterprise'', located in the provinces of Azuay, Morona Santiago y Cañar. They considered the period 2002--2012 and applied an Error Correction Model, getting as result a short-run price elasticity of 0.5383 and a long-run one of 0.2223. 

\protect\cite{i}, on other side, in their research determined---for the long-run---the ``Extra'' gasoline price and income elasticity as 0.4 and 0.32 respectively, and the ``Super'' gasoline ones as 1.55 and 0.6 respectively. They used monthly data regarding the period 1989--1998 and applied a Stock and Watson OLS dynamic approach. They compared it with results of Error Correction and Johansen models.

Finally, \protect\cite{j} analyzed the gasoline demand function of Ecuador for the period from 1979 to 2013 by applying a co-integration analysis and Error Correction Model, finding that the long-run income elasticity of ``Extra'' and ``Super'' gasoline is  0.202 and 0.251 respectively, suggesting that ``Super'' gasoline is a normal good in comparison with the ``Extra'' gasoline, which is inferior; moreover, she found no relationship between gasoline prices and gasoline demand. 

As seen above, though few researches had attempted to define price and/or income elasticities of the demand of specific kinds of energy in Ecuador, no one has aimed to state such parameters regarding energy demand at a global level.

\section{Background of energy demand in Ecuador} \label{background}

Ecuador, since its foundation, has been featured by keeping a growth model based mainly on the exportation of raw materials; however, its macroeconomic performance have been modified significantly overtime, so that its industry structure---essentially aimed to the satisfaction of its internal market---and beyond that, the behavior patterns of each economic sector within the country have also changed, affecting the evolution of the EC as well.  

During the first decades after the institutionalization of Ecuador as a Republic, the country's economic dynamic attempted to satisfy the demand of the international market with the production of goods in which the country presented ``comparative advantages'' so that---given the precariousness of the Ecuadorian industry at that time---the production of agricultural goods ended p being intensified and destined each time in greater proportions to the exportation. 

During the period 1860--1920, the ``cocoa boom'' took place in the ecuadorian economic context. The production of such a good used to be carried out in a rudimentary way which was intensive in respect to workforce. Its pick, however, finished due to an adverse context in the international market and the rise of a plague in the existent plantations around the 1930's. After that, from 1943 to 1965, another agricultural product---the banana---took the leadership among the ecuadorian exports and, beyond that, the boom of its exports developed along with the execution of a project aimed to lead the country to a structural industrial transformation, the ISI (Imports Substitution Industrialization) model, created and proposed by ECLAC (Economic Commission for Latin America and the Caribbean).  Such model sought to eliminate the importation of goods and to eradicate the agro-exporter model in order to encourage the modernization of the economy through the internal demand, so that this latter would become the creator of employment and added value.  

Until the 1970's, therefore, an agro-exporter growth model featured by two agricultural product booms took place in Ecuador. However, even though the model was set up in a still colonial context based on work in ``large states'', at the final of this period such a context had changed significantly due to a deeper introduction of the capitalist system in the economic structure and the development of different industrial sectors within the country. Furthermore, while at the beginning of the period people were dedicated either to a rudimentary agricultural production or to provide raw materials for the existent primitive manufacture in ``mitas'' and ``obrajes'', with the implementation of the ISI model, a greater level of technology was applied to the production: to the agricultural production as well as to the emerging industrial production related with food, beverages, tobacco and textiles. 

EC, consequently, also grew hastily during this period. According to the Economic Commission for Latin America and the Caribbean, in 1937 the EC in Ecuador was of about 761 millions of kwh (around 0.0654 Mtoe.); made up in 77\% of firewood, 16\% of oil and its derivatives and the rest of electricity (this latter used in 80\% by the industrial sector). Additionally, the oil production at that time was more than three times its consumption. Towards 1951, however, among other changes in EC, the oil production barely covered the consumption of oil and derivatives one time and a half. From 1929 to 1951, in fact, the total EC increased in about 800\% and the use of liquid fuels (which mostly comes from oil) increased due to its consumption by transportation, industry and residential sectors. 

In 1972, the petroleum mining and exportation at a big scale started to be developed---becoming in itself the next boom in the ecuadorian economy---along with an intensification of the industrialization process encouraged initially by the banana boom during the precedent decades. However, a series of events occurred during the next years such as the so-called ``debt crisis'' during the 80's and 90's decades---wherewith the ISI model was removed, the dolarization of the economy in the beginning of year 2000, the world financial crisis in 2008 and the current adverse international performance of oil prices had greatly affected such a process and, indeed, the level and structure of EC.   

\begin{figure}[ht]
    \centering
    \includegraphics[scale=0.6]{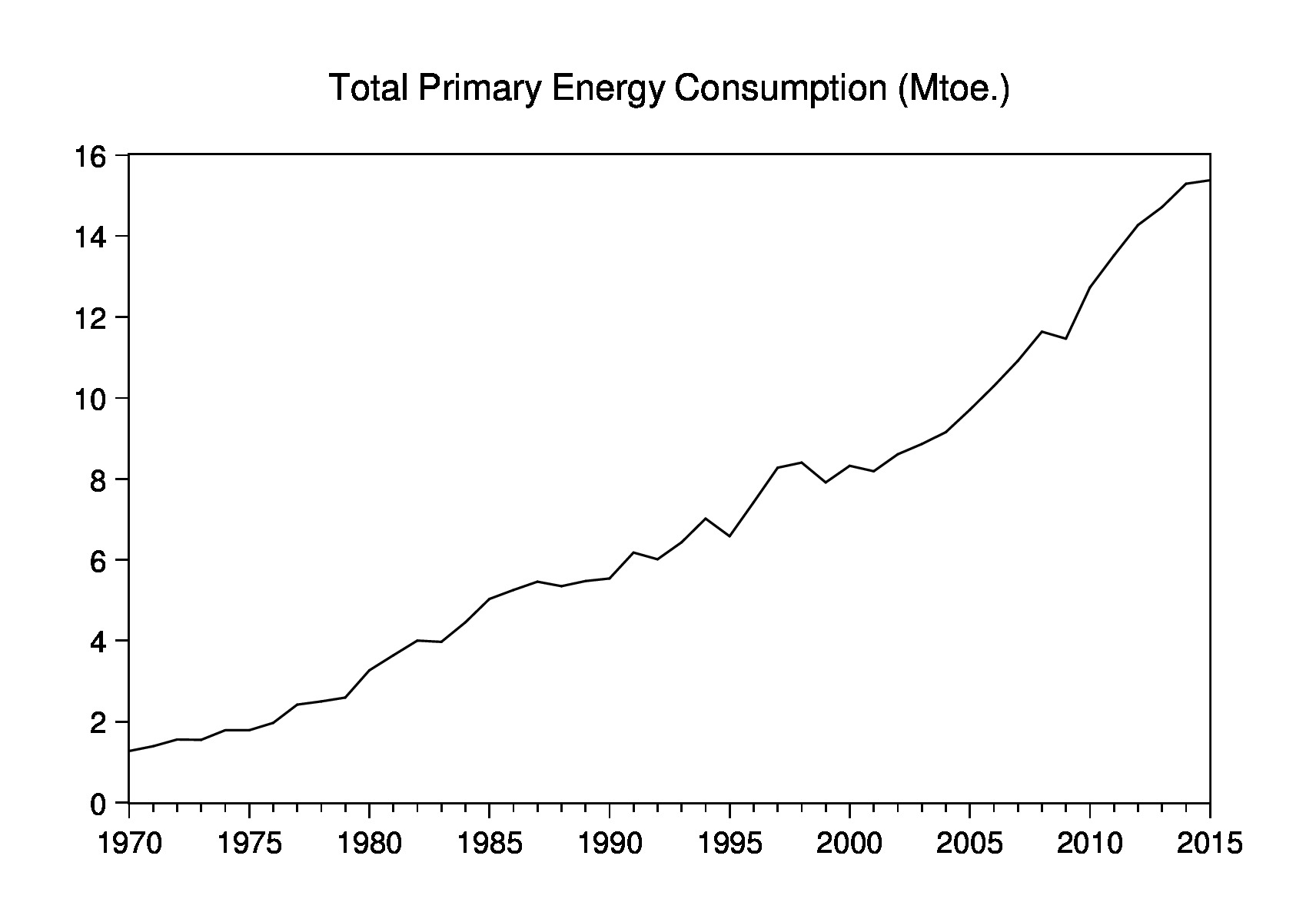}
    \caption{Evolution of Primary Energy Consumption}
    \label{Gpec}
\end{figure}

As shown in the Figure \ref{Gpec}, during the last 45 years the primary energy consumption (PEC)\footnote{Primary energy is every way of energy available in the nature before being converted or transformed, e.g.\ oil, coal, natural gas, solar energy, among others. Likewise, the secondary energy refers to any energy obtained from the transformation of primary energy, e.g.\ electricity, oil derivatives.} has had a general upward trend, showing a mean annual growth rate of 6\%, which was greater than the GDP's one (4\%, considering GDP in constant prices of 2007) in the same period. Moreover, during the last years the increasing in this variable has been even more accelerated. In fact, while from 1970 to 2000 the annual EC grew in 7.05516 Mtoe. (it passed from 1.27 to 8.32 Mtoe.), from 2000 to 2015 it increased in 7.05502 Mtoe. (passing from 8.32 to 15.38 Mtoe.), which shows that in the recent 15 years the annual EC has almost dobled, furthermore, in the period 2000--2015 it has increased in the same measure at which it increased from 1970 to 2000. 

\begin{figure}[ht]
    \centering
    \includegraphics[scale=0.6]{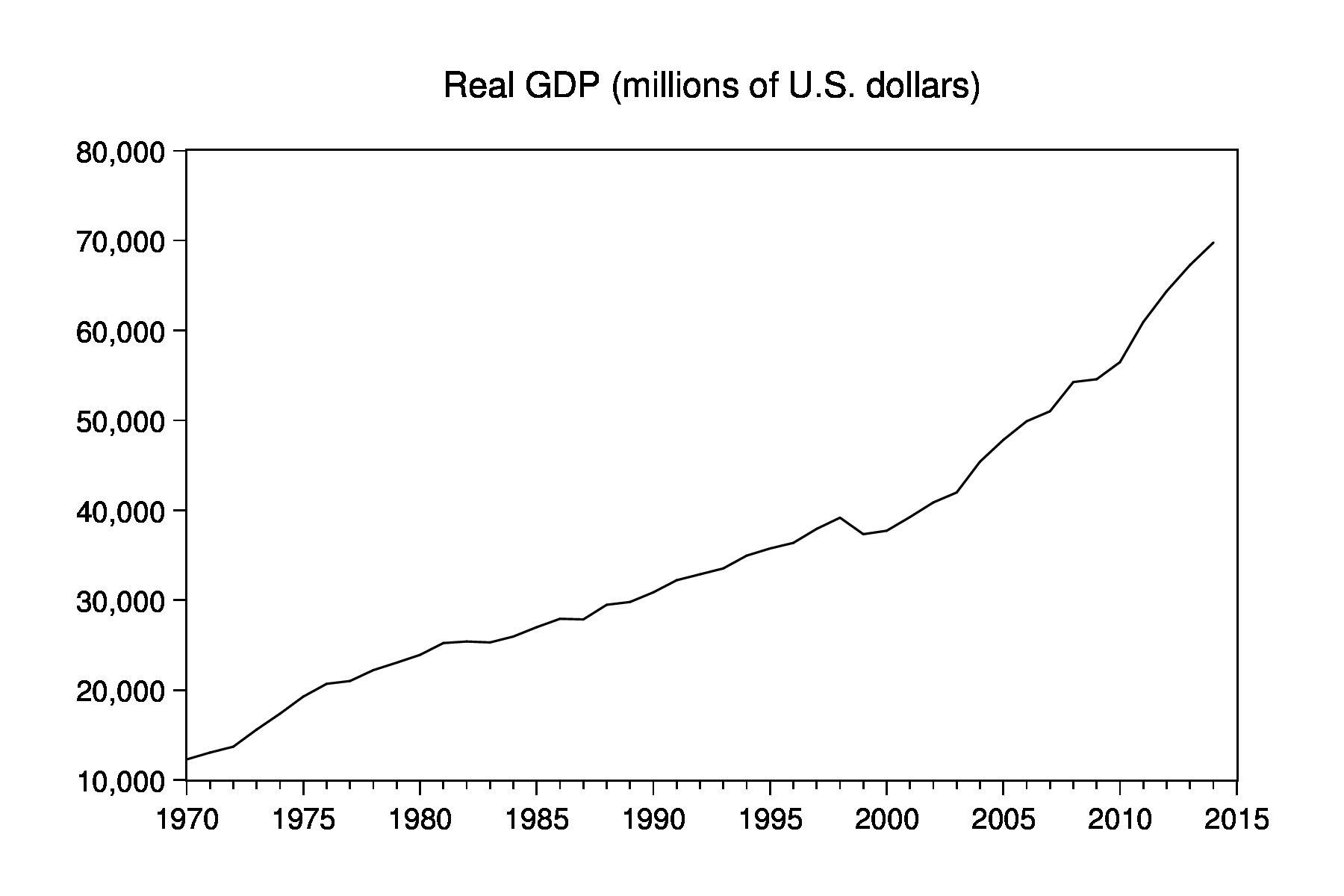}
    \caption{Evolution of real GDP}
    \label{Ggdp}
\end{figure}

Such considerable change, in fact, has been closely related with the evolution of ecuadorian GDP, especially during the last 15 years. As shown in the Figure \ref{Ggdp}, while before year 2000 the EC grew at a higher speed than GDP did, from 2000 to 2015 both of them have show a pretty similar trend, which agrees with the literature that establishes existent (either uni-directional---in both directions---or bi-directional) relationships between them. However, both variables have reflected the evolution of different economic sectors within the country and, indeed, the industrialization level and the consumption structure that such evolution has defined. 

Several important changes, in fact, have taken place in the ecuadorian economy in this context. According to information published by the Ministry Coordinator of Strategic Sectors of Ecuador, while in 1970 the primary energy was consumed in about 50\% by the residential sector, in 25\% by the transportation sector and in 10\% by industrial sector, in 2010 the the residential sector consumption was barely around 15\% of the total, proportion similar to which the industrial sector had, while the transportation sector embraced around the 50\%. From 2011 to 2014, furthermore, that scene hasn't changed significantly, so that in 2014 the transportation sector kept embracing 42\% of the total PEC. The transportation sector, therefore, has become the major energy consumer, while the proportion of EC by industrial sector has barely increased and residential consumption, by the other side, has decreased significantly (see Table \ref{TableECbyES}).

\begin{table}[htbp] 
\centering
\begin{threeparttable}
\footnotesize
  \caption{Structure of Ecuadorian Energy Consumption}
    \begin{tabular}{cccc}
    \toprule
    \multicolumn{4}{c}{\textbf{Energy Consumption by Economic Sector                           2011-2014}} \\
    \midrule
    \textbf{Year} & \textbf{Transportation Sector} & \textbf{Industrial Sector} & \textbf{Residential Sector} \\
    \midrule
    \textbf{2011} & 49.90\% & 9.40\% & 13.88\% \\
    \textbf{2014} & 42.00\% & 18.00\% & 12.00\% \\
    \bottomrule
    \bottomrule
    \end{tabular}
    \begin{tablenotes}
      \footnotesize
      \item Table prepared by the author. This table contains the proportions of energy consumption by the three major energy consumer sectors in Ecuador. 
      \item Information regarding 2011 was obtained from the presentation ``Energetic Matrix'', published by the Ministry Coordinator of Strategic Sectors of Ecuador. Regarding 2014, information was obtained from the ``National Energetic Balance 2015'', published by the same Ministry.
    \end{tablenotes}
  \label{TableECbyES}
  \end{threeparttable}
\end{table}

Additionally, oil products are the most used energy source by the transportation sector so that while in 2010 the participation of such products in its EC was  100\%, towards 2014 it barely decreased (it stood at 93.48\%). In general, during recent years the composition of EC by all the economic agents in Ecuador has been modified radically in respect to the scene of 1970: As mentioned in \protect\cite{CepalCE}, while in 1970 the total PEC was made up just in a marginal proportion by oil, in 2010 oil had a participation of about 75\%, while other sources of energy such as sugar cane, wood fire, hydro-energy and natural gas comprised the remaining 25\%. Undoubtedly, a greater participation of consumption of oil products has taken place. In fact, as shown in Table \ref{TableSEC}, on page \pageref{TableSEC}, while in 2011 the participation of such products in total EC stood at 78\%, in 2014 it stood at 76.56\%, remaining relatively constant. Moreover, although from 2011 to 2014 the consumption of oil products has decreased its participation in EC from transportation and residential sectors, it has increased its participation in the EC from industrial sector even at a greater rate than the diminishes mentioned. 

Additionally, according to the BP Statistical of World Energy 2016, in 2015 the total PEC was made up of 76.2\% by oil sources, 3.73\% by natural gas, 19.27\% by hidro-electricity and 0.8\% by renewable sources. Oil sources, consequently, keep being the mostly used in the country. Considering that, it becomes important to analyze how oil prices have evolved. 

\begin{figure}[ht]
    \centering
    \includegraphics[scale=0.6]{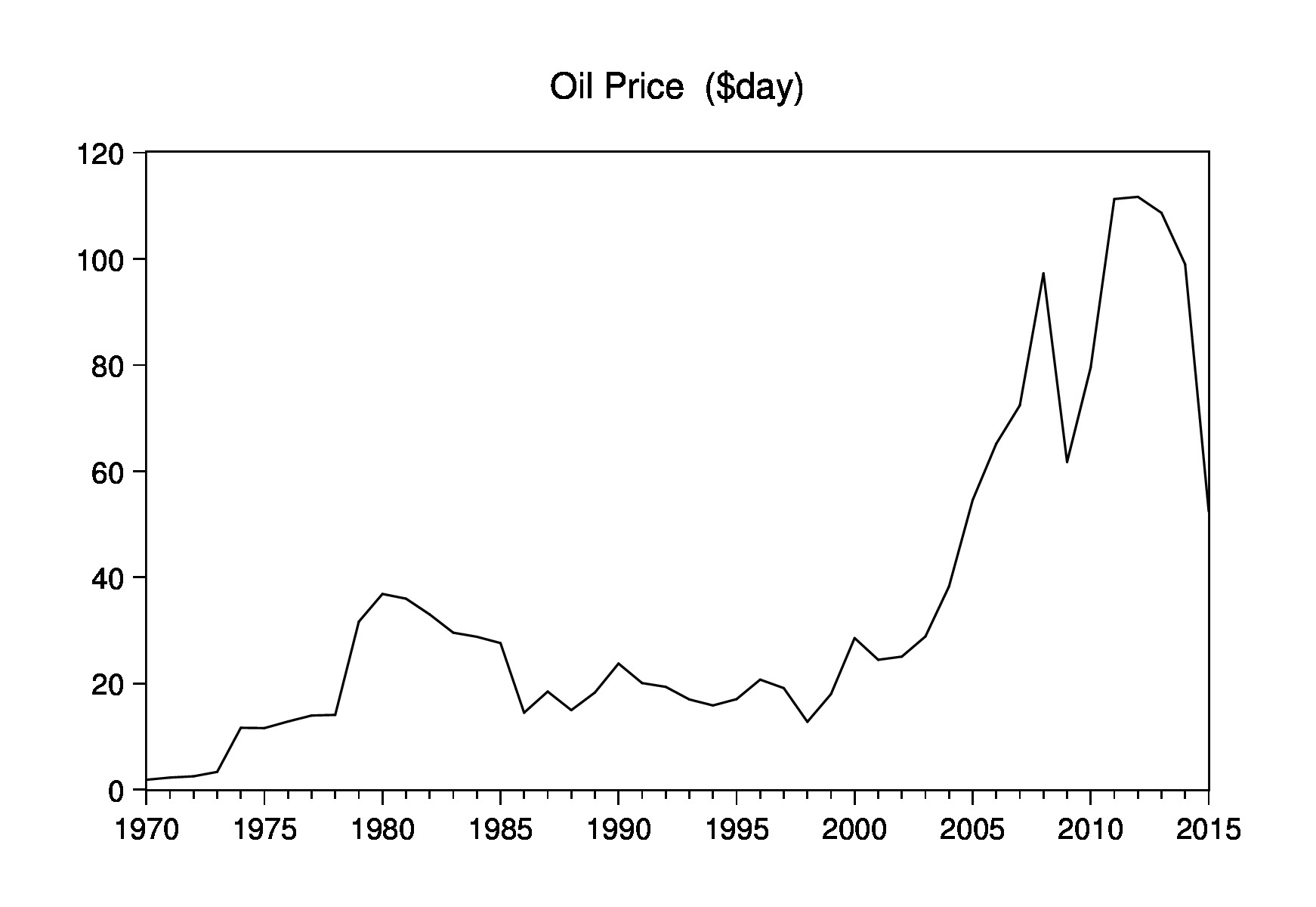}
    \caption{Evolution of International Oil Price}
    \label{OP}
\end{figure}

In general terms, the international oil price---considered as the ``Brent Oil" price since it is officially the reference for international oil price---evaluated in current U.S. dollars---and thus, the prices of oil derivatives---has had an increasing trend over time. Figure \ref{OP} shows the behavior of this series. It can be noticed that in 2009 (with the world financial crisis) oil price had a significant fall (it stood at \$61.67). Moreover, during the recent 3 years it has presented a decreasing trend: indeed, in 2015 it decreased 47.1\% in respect to the price of 2014.  

Even so, if we consider i) that Ecuador is a crude oil exporter country, ii) that approximately the 80\% of energy consumed corresponds to oil derivatives consumption, iii) that most of such derivatives (around 70\%, according to the Ministry Coordinator of Strategic Sectors of Ecuador) are imported and, iv) that the consumption of energy goods by all economic sectors is largely subsidized by the government, it is noticeable that the cost of acquiring energy goods is quite high in Ecuador. Moreover, the current increasing trend of EC with such a configuration is not sustainable in time since oil---whose exports have been able to fund such a level of EC---is a non-renewable energy source and furthermore, its price performance does not depend only on purely market factors but also responds to political issues. It turns important, therefore, carrying out research work which attempts to guide the policies  established by the government in the seeking of a more sustainable performance of the Ecuadorian economy.   

\section{Methodology and Data} \label{methodology}

\afterpage{
\begin{landscape}
\begin{table}[htbp]
\centering
\begin{threeparttable}
    \caption{Participation of energy sources in Energy Consumption by Economic Sector}
    \begin{tabular}{lcccccccr}
    \toprule
    \multicolumn{9}{c}{\textbf{COMPOSITION OF ENERGY CONSUMPTION 2011--2014}} \\
    \midrule
    \multicolumn{1}{c}{\multirow{2}[4]{*}{\textbf{Energy Sources}}} & \multicolumn{2}{c}{\textbf{Transportation Sector}} & \multicolumn{2}{c}{\textbf{Industrial Sector}} & \multicolumn{2}{c}{\textbf{Residential Sector}} & \multicolumn{2}{c}{\textbf{All Sectors*}} \\
\cmidrule{2-9}          & \textbf{2011} & \textbf{2014} & \textbf{2011} & \textbf{2014} & \textbf{2011} & \textbf{2014} & \textbf{2011} & \multicolumn{1}{c}{\textbf{2014}} \\
    \midrule
    \multicolumn{9}{l}{\textbf{Oil-based Energy}} \\
    Gasoline & 43.00\% & 40.52\% & 2.00\% & 0.03\% &       &       & 25.00\% & 28.20\% \\
    Diesel Oil & 42.00\% & 45.47\% & 15.00\% & 40.34\% &       &       & 32.00\% & 30.96\% \\
    Fuel Oil & 9.00\% & 7.29\% & 22.00\% & 13.92\% &       &       & 8.00\% & 8.08\% \\
    Jet Fuel & 6.00\% &       &       &       &       &       & 3.00\% &  \\
    GLP   &       & 0.20\% & 6.00\% & 3.39\% & 56.00\% & 52.70\% & 10.00\% & 8.52\% \\
    Others &       &       &       &       &       &       &       & 0.79\% \\
    \midrule
    \textbf{Total Oil-Based Energy} & \textbf{100.00\%} & \textbf{93.48\%} & \textbf{45.00\%} & \textbf{57.68\%} & \textbf{56.00\%} & \textbf{52.70\%} & \textbf{78.00\%} & \textbf{76.56\%} \\
    \midrule
    \midrule
    \multicolumn{9}{l}{\textbf{Non Oil-Based Energy}} \\
    Electricity &       & 0.01\% & 34.00\% & 28.28\% & 26.00\% & 33.10\% & 12.00\% & 13.56\% \\
    Firewood &       &       & 3.00\% &       & 18.00\% &       & 3.00\% & 2.00\% \\
    Gas Natural &       &       & 0.13\% & 1.44\% & 0.00\% &       & 0.01\% &  \\
    Productos de Caña &       &       & 14.00\% &       &       &       & 1.00\% & 1.91\% \\
    Others** &       & 6.51\% & 3.87\% & 12.60\% &       & 14.20\% & 5.99\% & 5.96\% \\
    \midrule
    \textbf{Total Non Oil-Based Energy} & \textbf{0.00\%} & \textbf{6.52\%} & \textbf{55.00\%} & \textbf{42.32\%} & \textbf{44.00\%} & \textbf{47.30\%} & \textbf{22.00\%} & \textbf{23.44\%} \\
    \bottomrule
    \bottomrule
    \end{tabular}
    \begin{tablenotes}
     \footnotesize
      \item Table prepared by the author. 
      \item Information regarding 2011 was obtained from the presentation ``Energetic Matrix'', published by the Ministry Coordinator of Strategic Sectors of Ecuador. Regarding 2014, information was obtained from the ``National Energetic Balance 2015'', published by the same Ministry.
      \item * Comprises, besides the specified sectors, the commerce, the agriculture, the fishing, the mining, the construction, the non-energetic sector, and the own consumption.
      \item ** Comprises kerosene, residuals, non-energetic materials and not specified.
      
    \end{tablenotes}
  \label{TableSEC}
  \end{threeparttable}
\end{table}
\end{landscape}
}

\subsection{Methodology} 

As mentioned by \protect\cite{c}, overtime the empirical methods used to investigate the determinants of energy demand---and, more specifically, its price and income elasticity---have changed substantially, but the modelling technique has generally remained the same: in the majority of studies energy demand model comprises a price variable and an income variable, regressed on some measure of EC.

Several investigations have considered just the GDP level and an Energy Price variable in the specification of Energy Demand equation as regressors, to estimate the income and price elasticities of energy demand, respectively \protect\citep[see, for example][]{a,b}.  \protect\cite{Jordan}, however, seeking to estimate such elasticities of energy demand in Jordan, along with an income and a price variable, included in the model the variable ``construction''---arguing that construction activity was a good indicator of the development process and indeed of energy usage---and a dummy to capture the changes in the political climate in Jordan. In the research of \protect\cite{c}, by the other hand, they looked for stating such elasticities including in the model the GDP level, energy price and a variable of energy efficiency as regressors; besides that, the authors used oil prices as energy prices arguing that energy used in Barbados was mostly from oil sources. Additional variables, however, have also been used in research work in order to define the determinants and elasticities of energy demand, such as the share of industry in GDP and the carbon dioxide emission levels \protect\citep[see, for instance][]{d}. 

Initially, the specification of energy demand model to be considered in this paper is as follows: 

\begin{equation} \label{eq:EC1}
\ln{E}_t=\ln{Y}_t+\ln{P}_t+\ln{I}_t+\varepsilon_t   
\end{equation}
where $E_t$ represents \textit{energy demand}, $Y_t$ represents the \textit{real income}, $P_t$ represents \textit{energy price}, and $I_t$ represents the \textit{industrial production} level. The model considers the series in logarithms since it is a  useful  practice for estimating  elasticities. The specific conceptualization of each variable included is as follows: $E_t$ is the per capita PEC, $Y_t$ is the real per capita GDP (at prices of 2007), $P_t$ , given that about the 80\% of EC is comprised by oil-based EC, is the international price of crude oil (valued in dollars of 2015), and $I_t$ is the Output of the Industrial Sector\footnote{Based on information of the World Bank. Industry includes manufacturing. It comprises value added in mining, manufacturing, construction, electricity, water, and gas. Value added is the net output of a sector after adding up all outputs and subtracting intermediate inputs. It is calculated without making deductions for depreciation of fabricated assets or depletion and degradation of natural resources.} (valued in prices of 2007).

The easiest and---by theoretical and empirical reasons---most desirable method to estimate the equation \eqref{eq:EC1} is the OLS method. However, as it is well known by the literature, when treating with time series, the OLS estimations are reliable just if the variables related in the model are stationary i.e\ they are I(0), otherwise such estimations would reflect no more than a spurious relationship between the variables. Therefore, tests for determining the  stationarity condition of each series must be applied. 

Some tests to seek for determining the integration level of variables, in  fact, have been developed to deal with such problem, mainly through the determination about the existence of unit roots in the series. Dickey Fuller (DF) and Augmented Dickey Fuller (ADF) tests, for instance, are some of the mostly used procedures; in such tests, the estimated statistic is compared to a set of critical values based in a null hypothesis about the existence of a unit a root i.e.\ the non-stationarity of the variables. Additionally, the Phillips and Perron test, is another of the mostly used tests, and corrects for any serial correlation and heteroskedasticity in the errors of the test regression by directly modifying the statistics estimated. One of the issues related to these tests which is being broadly analyzed and considered nowadays,  however, is the existence of points of structural break in the series, given that the existence of such points can wrongly influence unit root tests towards the acceptance of the null hypothesis of non-stationarity. 

\protect\cite{Perron88} developed a methodology to deal with this problem by the application of the ADF test on different specifications of the models of time series allowing for the existence of structural breaks. As mentioned in \protect\cite{URwithBExp}, this method suggests a general treatment of the structural break hypothesis
where four different situations are considered that allow a single break in the sample: (a) a change in the level, (b) a change in the level in the presence of a linear trend, (c) a change in the slope and (d) a change in both the level and slope.  In order to implement these models, two different transition mechanisms were considered: the additive outlier (AO) model where the transition is instantaneous and the trend break function is linear in parameters, and the innovation outlier (IO) model where changes occur via the innovation process and hence a gradual adjustment of a ``big'' shock takes place in accordance with the general dynamics of the underlying series.

The specification of the models considered by this method, regarding a given time series $y_t$, takes its start in a general model given as follows:

$y_t=\mu+\beta t+(1-\alpha L)^{-1}C(L)\varepsilon_t$.

Such model can also be expressed in the following way:

$\Delta y_t= (\alpha-1)y_{t-1}+\mu(1-\alpha)+\alpha\beta+(1-\alpha)\beta t + C(L)\varepsilon_t$,

where $(1-\alpha L)\mu_t=C(L)\varepsilon_t$, with $\mu_t$ such that $\varepsilon_t$ is i.i.d. and follows a normal distribution with mean 0 and variance $\sigma_\varepsilon^2$. $C(L)=\sum_{j=0}^{\infty}{c_j L^j}$, $\sum_{j=1}^{\infty}{j|c_j|}<\infty$, and $c_0=1$. $L$ is the lag operator. As mentioned by \protect\cite{URwithBExp}, it is important to note that the role of  deterministic component is different in the levels and the first differences representations. For instance, if $\alpha=1$, the constant term would equal to $\beta$ whereas the  slope would be 0. This shows the importance of carefully interpreting the meaning of deterministic terms under the null and alternative hypothesis.

The definition of the AO and IO models proposed by \protect\cite{Perron88} in order to test for the existence of unit root with a structural break, embraces the definition of two dummy variables: $DU_t$ and $DT_t$, such that $DU_t=1$ and $DT_t = t - T_1$ for $t>T_1$ and zero otherwise. $T_1$ represents the point of break in the series. 

According to the method, the following (AO) models are considered:

\begin{enumerate}
    \item $y_t=\mu_1+(\mu_2-\mu_1)DU_t+(1-\alpha L)^{-1}C(L)\varepsilon_t$
    \item $y_t=\mu_1+\beta t+(\mu_2-\mu_1)DU_t+(1-\alpha L)^{-1}C(L)\varepsilon_t$
    \item $y_t=\mu_1+\beta_1 t+(\beta_2-\beta_1)DT_t+(1-\alpha L)^{-1}C(L)\varepsilon_t$
    \item $y_t=\mu_1+\beta_1 t+(\mu_2-\mu_1)DU_t+(\beta_2-\beta_1)DT_t+(1-\alpha L)^{-1}C(L)\varepsilon_t$
\end{enumerate}

The first model considers a non-trending specification with a break in intercept, the second considers a trending specification with a break in intercept, the third one considers a trending specification with a break in trend and the fourth one considers a trending specification with a break in intercept and trend. 

The (IO) models considered are the following:

\begin{enumerate}
    \item $y_t=\mu+(1-\alpha L)^{-1}C(L)(\varepsilon_t+\theta DU_t)$
    \item $y_t=\mu+\beta t+(1-\alpha L)^{-1}C(L)(\varepsilon_t+\theta DU_t)$
    \item $y_t=\mu+\beta t+(1-\alpha L)^{-1}C(L)(\varepsilon_t+\theta DU_t+\gamma DT_t)$
\end{enumerate}

It  is  important to mention that these models consider known breakpoints. Additionally, as can be noticed, IO models do not embrace the case of a trending specification with break in trend, given that linear estimation methods cannot be used in this case as they are in the others. An alternative technique, however, was developed by \protect\cite{ZAur}, which provided a solution for the estimation of such model, at the time that developed a procedure  for  detecting break points endogenously from the  data. \protect\cite{ZAur} considered an estimated break point rather than a fixed one (an approach of endogenous break rather than one of an exogenous one). As \protect\cite{URthesis} mentions, the procedure to test for this kind of break is the same as for exogenous breaks: it tests for each possible break date in the sample, or some specific part of the sample, and then chooses the date with strongest evidence against the null hypothesis of unit root, i.e.\ where the t-statistic from the ADF test of unit root is  at a minimum\footnote{For a deeper explanation, see \protect\cite{ZAur}}. This method, however, does not allow for the existence of a break under the null hypothesis.

In order to test for  the existence of unit root considering two structural breaks, the methodologies developed by \protect\cite{Clem} and \protect\cite{LPur} are broadly used. \protect\cite{LPur} consider the presence of two breaks in trend variables, while \protect\cite{Clem} consider the existence of double change in mean. 

\protect\cite{LPur} consider (assuming that two breaks belong to the innovational outlier) the following model to carry out the test:

$\Delta y_t=\mu+\beta t+\theta DU1_t+\gamma DT1_t+\phi DU2_t+\Phi DT2_t+\alpha y_{t-1}+\sum_{i=1}^{k}{c_i \Delta y_{t-i}}+\varepsilon_t$, 

where $DU1_t=1$ and $DT1_t=t-TB1$ if $t>TB1$ and otherwise zero, $TB1$ is the first break point. $DU2_t=1$ and $DT2_t=t-TB2$ if $t>TB2$ and otherwise zero, $TB2$ is the second break point. The optimal lag length $(k)$ is determined based on the general to specific approach (the t test).

On the other hand, \protect\cite{Clem} consider (as \protect\cite{LPur}, assuming the case  of innovational outliers) the following model: 

$y_t=\mu+\rho y_{t-1}+\delta_1 DTB_{1t}+\delta_2 DTB_{2t}+d_1 DU_{1t}+d_2 DU_{2t}+\sum_{i=1}^{k}{c_i \Delta y_{t-i}}+\varepsilon_t$,

where $DTB_{it}$ is a pulse variable that takes the value 1 if $t=TB_i+1$ $(i=1,2)$ and 0 otherwise, $DU_{it}=1$ if $t<TB_i$ $(i=1,2)$ and 0 otherwise. $TB_i$ and $TB_2$ are the time periods when the mean is being modified. It is supposed that $TB_i=\lambda_i T$ $(i=1,2)$, with $0<\lambda_i<1$, and also that $\lambda_2>\lambda_1$. After such estimation, the procedure consists in obtaining the minimum value of the pseudo t-ratio for testing whether the autorregressive parameter is 1 for all the break time combinations. In order to derive the asymptotic distribution of the statistic, it is  assumed that $0<\lambda_0<\lambda_1$, $\lambda_2<1-\lambda_0<1$. Therefore, it is necessary to choose some trimming value ($\lambda_0$).  

After the determination about whether or not unit roots exist in the time series analyzed, and the nature of stationarity or non-stationarity of them, it is necessary to define exactly the procedure to follow in order to estimate the relationships among them, granting the estimation of a non-spurious but real relationship.  

In 1987, Engle and Granger faced the problem of the estimation of models when unit roots exist in the series by introducing the concept of ``cointegration'', establishing that, even being  individually non-stationary, if the variables are I(1) and there is a stationary linear combination of them, a correct estimation by OLS would be feasible through an Error Correction Model (ECM), which consists in the regression of the first difference of a dependent variable on its lags, lags of the first difference of regressors, and residuals of the OLS regression of variables in levels. This procedure, however, can potentially have a small sample bias and can examine at  most one cointegrating relationship between variables. 

The Johansen's Vector Error Correction Model (VECM), developed in 1988, for instance, presents several advantages over the ECM: (1) it does not assume one co-integrating relationship, (2) it does not impose any exogeneity restrictions and (3) it uses a system of equations framework to estimate the model. This methodology takes its starting point in the vector auto regression (VAR) of order $p$ given by 

$\M{X}_t= \M{\Phi}_0+\M{\Phi}_1\M{X}_{t-1}+\ldots+\M{\Phi}_p\M{X}_{t-p}+\M{\varepsilon}_t$,

where $\M{X}$ is a $(n\times1)$ vector of variables that are integrated of order one i.e.\ I(1) and $\M{\varepsilon}_t$ is a $(n\times1)$ vector of innovations. This VAR can also be written as 

$\Delta\M{X}_t= \M{\Phi}_0+\M{\Pi}\M{X}_{t-1}+\sum_{i=1}^{p-1}{\Gamma_i{\Delta\M{X}}_{t-i}}+\M{\varepsilon}_t$,

where it holds that

$\M{\Pi}=\sum_{i=1}^{p}{\M{\Phi}_i}-\M{I}$

$\M{\Gamma}_i=-\sum_{j=i+1}^{p}{\M{\Phi}_j}$, $j = 1, \ldots,p-1$

Besides, $r$ is the range of the matrix $\M{\Pi}$. If $r=0$, then there is no co-integration, so that non-stationarity of $I(1)$ type in the variables vanishes by taking differences. If $r=n$ , this is, if $\M{\Pi}$ has full rank, then the variables of $\M{X}$ cannot be $I(1)$ but are stationary, so that analysis with variables in level could be carried out. Finally, if $\M{\Pi}$ has reduced rank, this is, if $0<r<n$, then at least one cointegrating relationship exists and $r$ represents the number of cointegrating relationships. 

Furthermore, Johansen proposes two different likelihood ratio tests of the significance of the reduced rank of the $\M{\Pi}$ matrix, in order to determine the number of existent cointegrating relationships between the variables: the trace test and maximum eigenvalue test, shown in equations: 

$J_{\text{trace}}=-T\sum_{i=r+1}^{n}{\ln(1-\hat{\lambda}_i)}$

$J_{\max}=-T\ln(1-\hat{\lambda}_i)$

Here, $T$ is the sample size and $\hat{\lambda}_i$ is the $i$th largest canonical correlation of $\Delta \M{X}_t$ with $\M{X}_{t-1}$. The trace test tests the null hypothesis of $r$ cointegrating vectors against the alternative hypothesis of $n$ cointegrating vectors.  The maximum eigenvalue test, on the other hand, tests the null hypothesis of $r$ cointegrating vectors against the alternative hypothesis of $r+1$ co-integrating vectors.  Neither of these test statistics follows a chi square distribution. Asymptotic critical values for this procedure were provided by \protect\cite{ostercv} and \protect\cite{mackcv}. 

As many authors mention, however, results of both procedures are completely reliable in large samples. For the case of small samples, therefore, an alternative procedure has been developed and is widely used nowadays: the Dynamic OLS (DOLS) approach. This method was proposed  by \protect\cite{SW} and basically is an improvement of OLS estimation by dealing with small sample and dynamic sources of bias. As \protect\cite{Jordan} mentions, it is a robust single equation approach which corrects for regressor endogeneity by the inclusion of leads and lags of first differences of the regressors, and for serially correlated errors by a GLS procedure. This method allows to estimate the long-run equilibrium, in systems which may involve variables either integrated of the same order or integrated of different orders but still cointegrated. In addition, it has the same optimality properties as the Johansen distribution. In this paper, essentially due to its robustness in small samples in comparison with precedent methods, the DOLS approach is applied. 

The DOLS model, in general terms and for a dependent variable $Y_t$ with regressors $X_{i,t}$, $i=1,2, \ldots, k$, is given as follows:

$Y_t= \beta_0 + \beta_1X_{1,t}+\beta_2X_{2,t}+\ldots+\beta_kX_{k,t}+\sum_{i=p}^{q}{\alpha_i{\Delta X_{1,t-i}}}+\sum_{i=p}^{q}{\delta_i{\Delta X_{2,t-i}}}+\ldots+\sum_{i=p}^{q}{\phi_i{\Delta X_{k,t-i}}}+\varepsilon_t$

where $p$ and $q$ represent the maximum orders of leads and lags (respectively) of the first differences of the regressors included in the model. 

\subsection{Data}

All time series considered contain annual information spanning the period 1970--2015. The information about the real per capita GDP and the Total Output of Industry in Ecuador was extracted from the website of the World Bank. The information about the PEC level and the International Crude Oil Prices (in U.S. dollars of 2015) are from the BP Statistical Review of World Energy 2016, published in the website of BP Global.

\section{Results} \label{results}

Before moving on to the estimation of equation \ref{eq:EC1}, it is important to test for the existence of unit roots in each series involved. Firstly, ADF and Phillips-Perron tests are  applied to the series. The specification of the model for testing the existence of unit root (intercept and trend, intercept, and none) is chosen considering the significance of the regressors included in the model. The results obtained can be seen in the Table \ref{Results1}.  In all cases (except in the case of $\ln{P}$) the regressors considered in the model of the tests are significant at 95\%. In fact, even the test specification embracing neither constant nor trend (shown in Table \ref{Results1}) is found non-significant for the case of $\ln{P}$. The results suggest that $\ln{E}$ is  stationary without drift, $\ln{Y}$  is  non-stationary and $\ln{I}$ is stationary around a trend at 95\% level. No concrete conclusion, however, is obtained about $\ln{P}$.

\begin{center}
\begin{table}[htbp]
 \centering
  \begin{threeparttable}
  \caption{Results of ADF and PP Tests}
\footnotesize
    \begin{tabular}{lccccccc}
    \toprule
    \multicolumn{1}{c}{\multirow{2}[4]{*}{\textbf{TEST}}} & \multirow{2}[4]{*}{\textbf{Model Specification}} & \multirow{2}[4]{*}{\textbf{Statistic}} & \multicolumn{3}{c}{\textbf{Critical Values}} & \multirow{2}[4]{*}{\textbf{Lags}} & \multirow{2}[4]{*}{\textbf{Bandwidth}} \\
\cmidrule{4-6}          &       &       & \textbf{1\%} & \textbf{5\%} & \textbf{10\%} &       &  \\
    \midrule
    \textbf{$\ln{Y}$} &       &       &       &       &       &       &  \\
    ADF   & No Constant, No Trend & 3.700 & -2.617 & -1.948 & -1.612 & 0     &  \\
    PP    & No Constant, No Trend & 2.617 & -2.617 & -1.948 & -1.612 &       & 4 \\
    \midrule
    \textbf{$\ln{P}$*} &       &       &       &       &       &       &  \\
    ADF   & No Constant, No Trend & 0.348 & -2.617 & -1.948 & -1.612 & 0     &  \\
    PP    & No Constant, No Trend & 0.346 & -2.617 & -1.948 & -1.612 &       & 2 \\
    \midrule
    \textbf{$\ln{E}$} &       &       &       &       &       &       &  \\
    ADF   & No Constant, No Trend & -4.508 & -2.617 & -1.948 & -1.612 & 0     &  \\
    PP    & No Constant, No Trend & -4.946 & -2.617 & -1.948 & -1.612 &       & 1 \\
    \midrule
    \textbf{$\ln{I}$} &       &       &       &       &       &       &  \\
    ADF   & Trend and Constant & -3.835 & -4.176 & -3.513 & -3.187 & 0     &  \\
    PP    & Trend and Constant & -3.734 & -4.176 & -3.513 & -3.187 &       & 4 \\
    \bottomrule
    \end{tabular}
    \begin{tablenotes}
      \footnotesize
      \item Akaike Information Criterion was considered to choose the appropriate number of lags in ADF test. 
      \item Bartlett  kernel spectral estimation method and Newey-West bandwidth  selection were considered in PP test.
    \end{tablenotes}
  \label{Results1}
\end{threeparttable}
\end{table}
\end{center}

\begin{figure}[ht]
    \centering
    \includegraphics[scale=0.6]{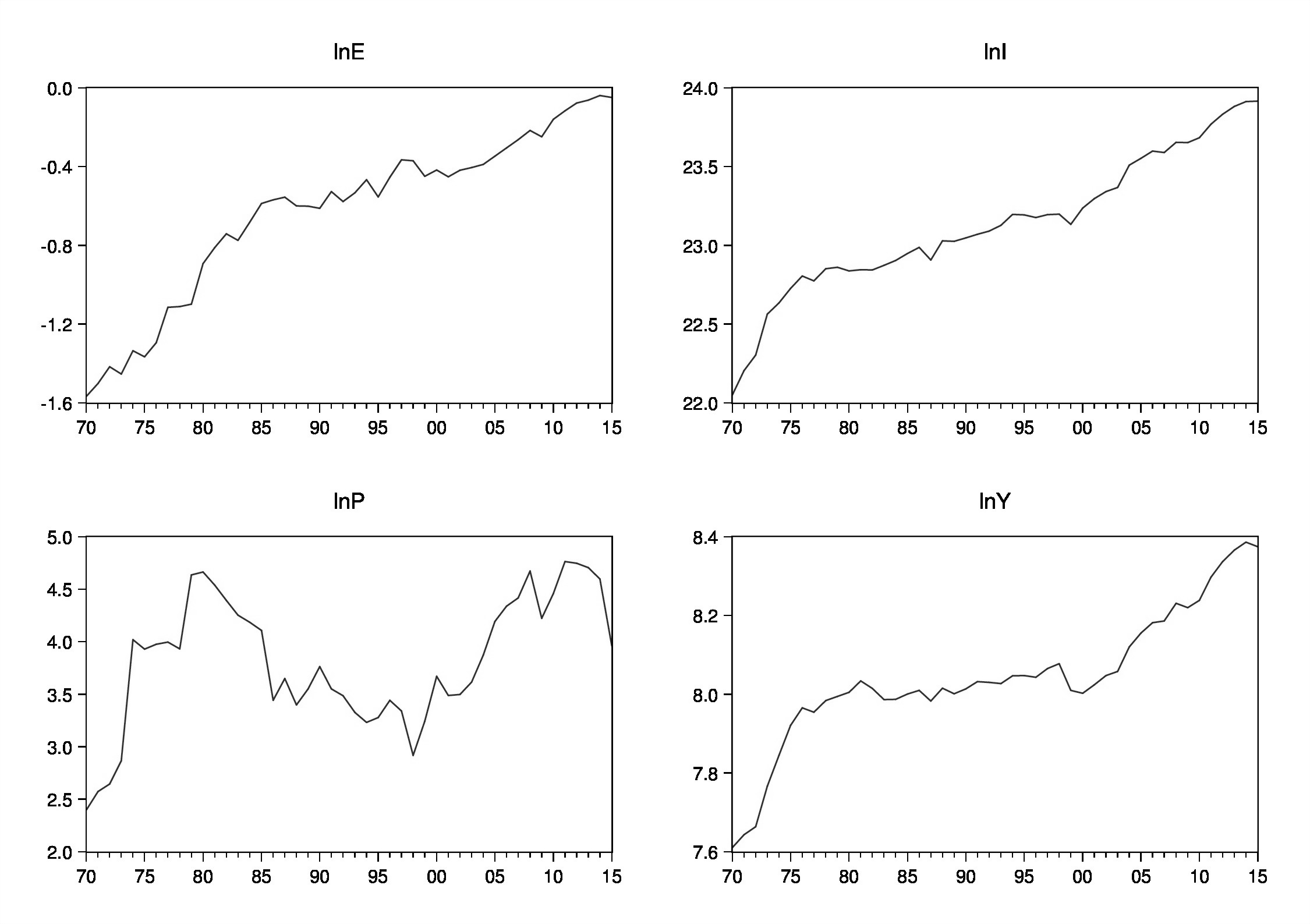}
    \caption{Behavior of Time Series Considered}
    \label{4graphs}
\end{figure}

Therefore, as the next step, unit root tests allowing for structural breaks are carried out. The consideration of structural breaks (either in intercept or in trend) is  important particularly in  this analysis since during the period of study several facts have affected significantly the economic performance of Ecuador. The behavior of the series considered can be appreciated in the Graph \ref{4graphs}. Since the series suggest to present gradual changes, the type of break to be considered in all cases is ``Innovational Outlier'' (IO). 

The testing for unit roots in consideration of structural breaks starts by considering only one break point in each series. The specification of the model for testing the existence of unit root (intercept and trend, intercept, and none) as well as the specification of the break included in it (break in intercept, in trend, or in both) is chosen considering the significance of the regressors included in the model. The Table \ref{Results2} shows the results obtained. The whole specification considered (of the deterministic component as well as of the break) regarding the series $\ln{E}$, $\ln{Y}$ and $\ln{I}$, is significant at  95\% of confidence. In the case of $\ln{P}$, even the specification without considering either trend or intercept (shown in the Table \ref{Results2}), is not significant.  

\begin{center}

\begin{table}[htbp]
  \centering
  \begin{threeparttable}
  \caption{Results of Unit Root Tests considering One Structural Break}
\footnotesize 
    \begin{tabular}{cccccc}
    \toprule
    \multirow{2}[4]{*}{\textbf{Variable}} & \multicolumn{2}{c}{\textbf{Model Specification}} & \multirow{2}[4]{*}{\textbf{t-Statistic}} & \multirow{2}[4]{*}{\textbf{Lags}} & \multirow{2}[4]{*}{\textbf{Break}} \\
\cmidrule{2-3}          & \textbf{Deterministic Component} & \textbf{Break } &       &       &  \\
    \midrule
    \textbf{$\ln{E}$} & Constant and Trend & Trend  & -4.207 & 0     & 1983 \\
    \midrule
    \textbf{$\ln{Y}$} & Constant and Trend & Trend  & -4.942 & 4     & 2004 \\
    \midrule
    \textbf{$\ln{P}$*} & Constant & Constant  & -3.202 & 0     & 1998 \\
    \midrule
    \textbf{$\ln{I}$} & Constant and Trend & Trend  & -5.308 & 0     & 2000 \\
    \bottomrule
    \end{tabular}
    \begin{tablenotes}
    \footnotesize
    \item The procedure developed in \protect\cite{Perron88} and \protect\cite{VPerron98} was considered for the test of $\ln{P}$. The procedure of \protect\cite{ZAur} was considered in tests of $\ln{E}$, $\ln{Y}$ and $\ln{I}$. 
    \item The critical values at 95\% of confidence are -4.52 for series $\ln{E}$, $\ln{Y}$ and $\ln{I}$, and -4.44 for $\ln{P}$.
    \item The number of lags selected in each case was chosen based on Akaike  Information Criterion  (AIC).
    \end{tablenotes}
  \label{Results2}
\end{threeparttable}
\end{table}
\end{center}

Thus, based  on such results, regarding series $\ln{E}$, $\ln{Y}$ and $\ln{I}$, it could be concluded that: (1) $\ln{Y}$, as well as $\ln{I}$ are trend stationary variables, with structural (innovative) breaks in years 2004 and 2000, respectively and (2) $\ln{E}$ is a non-stationary variable. In addition, conclusive results about the existence of unit root in $\ln{P}$ were not found.

Up to here, only $\ln{Y}$  and $\ln{I}$ series have a specifically defined stationarity condition based in the tests carried out, whereas $\ln{E}$ and $\ln{P}$ have been found to be non-stationary without any other specification. In order to obtain a better understanding about the stationarity condition of $\ln{P}$  and $\ln{E}$, tests of unit roots considering two break points were performed regarding this series. The test of \protect\cite{LPur} was performed firstly, the results are shown in Table \ref{Results3}. As not all the dummies included in each model were significant, results suggest that both series do not show a trending behavior even when considering 2 breaks. Therefore, the test of \protect\cite{Clem} is carried out about these series. The results can be seen in Table \ref{Results4} and suggest that $\ln{E}$ and $\ln{P}$ are non stationary variables; however, $\ln{E}$ presents significant structural breaks in the years 1975 and 2004, while $\ln{P}$ presents them in years 1984 and 2002.  

\begin{center}
\begin{table}[htbp]
  \centering
  \begin{threeparttable}
  \caption{Results of Lumsdaine and Papell Tests}
  \footnotesize
    \begin{tabular}{cclcccc}
    \toprule
    \multicolumn{3}{c}{\textbf{Dummy Variables}} & \multirow{2}[4]{*}{$\alpha$} & \multirow{2}[4]{*}{\textbf{t-Statistic}} & \multirow{2}[4]{*}{\textbf{Number of  Lags}} & \multirow{2}[4]{*}{\textbf{Break Points}} \\
\cmidrule{1-3}    \textbf{Variable} & \textbf{Coef.} & \multicolumn{1}{c}{\textbf{Std. Error}} &       &       &       &  \\
    \midrule
    \midrule
          &       &       &       &       &       &  \\
    \multicolumn{1}{l}{\textbf{$\ln{E}$}} &       &       &       &       &       &  \\
    DU1   & 0.114 & 0.062 & \multirow{4}[1]{*}{-0.503} & \multirow{4}[1]{*}{-5.050} & \multirow{4}[1]{*}{0} & \multirow{4}[1]{*}{1976  / 1980} \\
    DT1   & 0.012 & 0.023 &       &       &       &  \\
    DU2   & 0.147 & 0.048** &       &       &       &  \\
    DT2   & -0.002 & 0.022 &       &       &       &  \\
    \midrule
          &       &       &       &       &       &  \\
    \multicolumn{1}{l}{\textbf{$\ln{P}$}} &       &       &       &       &       &  \\
    DU1   & 0.565 & 0.268** & \multirow{4}[1]{*}{-0.330} & \multirow{4}[1]{*}{-3.526} & \multirow{4}[1]{*}{0} & \multirow{4}[1]{*}{1979 / 1986} \\
    DT1   & -0.095 & 0.051 &       &       &       &  \\
    DU2   & -0.195 & 0.238 &       &       &       &  \\
    DT2   & 0.112 & 0.051** &       &       &       &  \\
    \bottomrule
    \end{tabular}
    \begin{tablenotes}
    \footnotesize
    \item Highlighted standard errors denote variables significant (at 95\% of confidence).
    \item Number of lags included were determined based on the general to specific approach (the t test).
    \item For all series, the trimming parameter was established equal to 0.1.
    \end{tablenotes}
  \label{Results3}
\end{threeparttable}
\end{table}
\end{center}

\begin{center}
\begin{table}[htbp]
  \centering
  \begin{threeparttable}
  \caption{Results of Clemente-Montañez-Reyes Tests}
  \footnotesize
    \begin{tabular}{ccccccc}
    \toprule
    \multicolumn{3}{c}{\textbf{Dummy Variables}} & \multirow{2}[4]{*}{$\rho -1$} & \multirow{2}[4]{*}{\textbf{t-Statistic}} & \multirow{2}[4]{*}{\textbf{Number of  Lags}} & \multirow{2}[4]{*}{\textbf{Break Points}} \\
\cmidrule{1-3}    \textbf{Variable} & \textbf{Coef.} & \textbf{Std. Error} &       &       &       &  \\
    \midrule
    \midrule
          &       &       &       &       &       &  \\
    \multicolumn{1}{l}{\textbf{LNE}} &       &       &       &       &       &  \\
    DU1   & 0.13207 & 3.156*** & -0.1793 & -4.427 & 1     & 1975  / 2004 \\
    DU2   & 0.08032 & 2.983*** &       &       &       &  \\
    \midrule
          &       &       &       &       &       &  \\
    \multicolumn{1}{l}{\textbf{LNP}} &       &       &       &       &       &  \\
    DU1   & -0.73789 & -4.669*** & -0.6147 & -5.139 & 4     & 1984 / 2002 \\
    DU2   & 0.77171 & 5.053*** &       &       &       &  \\
    \bottomrule
    \end{tabular}
    \begin{tablenotes}
    \footnotesize
    \item Highlighted standard errors denote variables which are significant at 95\% of confidence. 
    \item The critical value at 95\% of confidence is -5.490. 
    \end{tablenotes}
  \label{Results4}
  \end{threeparttable}
\end{table}
\end{center}

What we know at this point is (1) that $\ln{Y}$ and $\ln{I}$ were found to be trend stationary with structural breaks in years 2004 and 2000, respectively. (2) $\ln{E}$ is  a non-stationary variable  with significant breaks in mean in 1975 and 2004, and (3) $\ln{P}$ is a non-stationary variable with significant breaks in mean in years 1984 and 2002.  The next step is, therefore, testing for the existence of cointegration relationships between the variables in presence of the deterministic components and structural breaks found.

Through the tests applied regarding the existence of unit roots in the series, structural breaks in them were found for years 1975, 1984, 2000, 2002 and 2004. Year 1975 is near to 1973, year at which officially started the oil prices crisis of 70's. Year 1984, is near to 1983, year at which officially started  the so-called debt crisis of 80's in Ecuador (in this year, Ecuador signed its first Intention Letter). In addition, 2000, 2002 and 2004 are years that refers essentially to year 2000, since in this year the debt crisis in the country gave way officially to a dolarization process and so, to a change in Ecuadorian economic performance. Only two possible breaks are going to be considered in the modelling energy demand, years 1983 and 2000, since a break in 1973 could represent a problem given the period of data considered in the study.

In order to define the number of lags to include in testing for cointegration (through the Johansen's method), VAR  models were fitted. The exogenous variables considered in such VAR models were: (1) dummies of the breaks found ($B_1$ and $B_2$ regarding  year 1983 and 2000 respectively), (2) a trend variable ($T$) and (3) interactions between (1) and (2). Regarding the breaks included in VAR models, 3 different break dummy specifications are assessed: (1) 1983  (2) 2000 and  (3) 1983--2000. Schwarz, Akaike and Hannan-Quinn Information Criterion were considered in choosing the adequate number of lags. Thereafter, Johansen's cointegration test was performed considering the series. Asymptotic critical values developed by \protect\cite{giles} for the test  of cointegration  in presence of structural   breaks, were calculated.

\afterpage{
\begin{landscape}
\vspace*{2.8cm}
\begin{table}[htbp]
  \centering
  \begin{threeparttable}
    \caption{Results of Johansen's Cointegration Tests in consideration of Structural Breaks}
    \begin{tabular}{ccccccccc}
    \toprule
    \multirow{2}[4]{*}{\textbf{Break}} & \multirow{2}[4]{*}{\textbf{Lags}} & \multirow{2}[4]{*}{\textbf{Exogenous Variables}} & \multirow{2}[4]{*}{\textbf{Range of $\M{\Pi}$}} & \multirow{2}[4]{*}{\textbf{Trace Statistic}} & \multicolumn{3}{c}{\textbf{Asymptotic Critical Values}} & \multirow{2}[4]{*}{\textbf{Conclusion}} \\
\cmidrule{6-8}          &       &       &       &       & \textbf{90\%} & \textbf{95\%} & \textbf{99\%} &  \\
    \midrule
    \multirow{4}[2]{*}{1983} & \multirow{4}[2]{*}{1} & \multirow{4}[2]{*}{$B_{1}$, $T(B_{1})$} & $r=0$   & 78.78 & 78.38 & 82.60 & 90.91 & Rejection  \\
          &       &       & $r \leq 1$  & 44.03 & 53.85 & 57.43 & 64.57 & Acceptance  \\
          &       &       & $r \leq 2$  & 17.87 & 33.13 & 36.06 & 41.98 & ----- \\
          &       &       & $r \leq 3$  & 6.09  & 16.05 & 18.24 & 22.85 & ----- \\
    \midrule
    \multirow{4}[2]{*}{2000} & \multirow{4}[2]{*}{1} & \multirow{4}[2]{*}{$B_2$, $T(B_2)$} & $r=0$   & 105.51 & 78.95 & 83.18 & 91.52 & Rejection  \\
          &       &       & $r \leq 1$  & 67.04 & 54.35 & 57.95 & 65.11 & Rejection  \\
          &       &       & $r \leq 2$  & 33.34 & 33.53 & 36.47 & 42.40 & Acceptance  \\
          &       &       & $r \leq 3$  & 9.16  & 16.28 & 18.46 & 23.04 & ----- \\
    \midrule
    \multirow{4}[2]{*}{1983, 2000} & \multirow{4}[2]{*}{1} &       & $r=0$   & 105.24 & 100.57 & 105.37 & 114.77 & Rejection  \\
          &       & $B_{1}$, $T(B_{1})$  & $r \leq 1$  & 59.29 & 71.13 & 75.23 & 83.34 & Acceptance  \\
          &       & $B_2$, $T(B_2)$ & $r \leq 2$  & 28.99 & 45.34 & 48.69 & 55.38 & ----- \\
          &       &       & $r \leq 3$  & 8.42  & 22.62 & 25.06 & 30.07 & ----- \\
    \bottomrule
    \end{tabular}
    \begin{tablenotes}
    \footnotesize
    \item Number of lags in each case was defined by running VAR models considering structural breaks as exogenous variables. 
    \item Critical values were calculated according to \protect\cite{giles}.
    \item Acceptance  and rejection defined in all cases at 90\% of confidence.
    \end{tablenotes}
    \label{Results5}
    \end{threeparttable}
\end{table}
\end{landscape}
}

As illustrated by  Table \ref{Results5}, on page \pageref{Results5}, the results suggest that at least one cointegrating relationship exists between the variables when considering structural breaks (1) in year  1983 and (2) in years 1983 and 2000. Regarding year 2000, according to the results, 2 cointegration relationships exist. However, since the retained methodology is DOLS (because of  its reliability according to  the sample size), this option is not considered. Thus, only the models with breaks (1) in year 1983 and (2) in years 1983 and 2000 are estimated. It has been selected only one lag and no lead for such estimations. The Dynamic OLS models to be fitted, therefore, are the following:

Model 1:~
\begin{align*}
\ln E_t & =\beta_0+\beta_1\ln Y_t+\beta_2\ln P_t+\beta_3\ln I_t \\ 
&  +\gamma_1 T+\gamma_2 B_{1,t}+\gamma_3 T(B_{1,t}) \\
&  +\sum_{i=0}^{1}{\alpha_i \Delta \ln{Y}_{t-i}}+\sum_{i=0}^{1}{\delta_i \Delta \ln{P}_{t-i}}+\sum_{i=0}^{1}{\phi_i \Delta \ln{I}_{t-i}}+\varepsilon_t
\end{align*}

Model 2:~
\begin{align*}
\ln E_t & =\beta_0+\beta_1\ln{Y}_t+\beta_2\ln{P}_t+\beta_3\ln{I}_t \\
& +\gamma_1 T+\gamma_2 B_{1,t}+\gamma_3 T(B_{1,t})+\gamma_4 B_{2,t}+\gamma_5 T(B_{2,t})\\
&  +\sum_{i=0}^{1}{\alpha_i \Delta \ln{Y}_{t-i}}+\sum_{i=0}^{1}{\delta_i \Delta \ln{P}_{t-i}}+\sum_{i=0}^{1}{\phi_i \Delta \ln{I}_{t-i}}+\varepsilon_t
\end{align*}
where $B_{1}$ and $B_{2}$ are dummy variables that represent the chosen years of structural break (1983 and 2000, respectively), and $T$ is the trend variable. Clearly, the models to be fitted estimate the long-run relationship between $\ln{E}$ and its determinants ($\ln{Y}$, $\ln{P}$ and $\ln{I}$), by taking into account the lagged effect of those 3 variables as well as the effect of structural breaks in the series (either in intercept or in trend). The results about the estimation of the models is shown in Table \ref{Results6}. As can be noticed, $R^2$, the standard error of the regression and the long-run variances of both models suggest that Model 1 and 2  are good representations of the dynamics between the variables. The Jarque-Bera statistic in both models suggests that residuals follow a normal distribution at a 5\% of significance. Additionally, according to the respective statistics shown in Table \ref{Results6}, the significance of $\ln{Y}$, $\ln{P}$ and $\ln{I}$ as regressors in Models 1 and 2 varies across models, however, the variable $\ln{P}$ is not significant in any model. Since in Model 1 variables are significant at a better level than in model 2, results shown regarding Model 1 are considered as a better estimation.

Thereby, based on results obtained in Model 1, the following can be concluded: At the long run, (1) the income elasticity of energy demand is 1.8, which (since it is greater than 1) allows to state that energy demand is highly income elastic.   This result agrees with the close relationship observed graphically among the trends of the GDP and EC overtime. And (2) energy demand presents an almost unitary elasticity (-1.2) in relation with the production level of industrial sector, however, such significant relationship is inverse. This result can be explained by the evolution of EC from the economic sectors within the country: Keeping constant the level of investment in the country, an increasing in the total industrial output might imply the movement of economic factors from other economic sectors---such as the transportation sector, which is the major energy consumer---towards the industrial one, causing a slight decrease rather than an increase in the total level of EC.

\begin{center}
\begin{table}[htbp]
  \centering
  \begin{threeparttable}
  \caption{Results of Estimation of Model 1 and Model 2}

    \begin{tabular}{crrrr}
    \toprule
    \textbf{Variable} & \multicolumn{2}{c}{\textbf{Model 1}} & \multicolumn{2}{c}{\textbf{Model 2}} \\
    \midrule
          &       &       &       &  \\
    $\ln{Y}$   & 1.765902 &       & 1.225363 &  \\
          & [0.330706] & \multicolumn{1}{l}{***} & [0.580153] & \multicolumn{1}{l}{**} \\
    $\ln{P}$   & 0.006475 &       & -0.027041 &  \\
          & [0.024808] &       & [0.041382] &  \\
    $\ln{I}$   & -1.219197 &       & -1.017343 &  \\
          & [0.222253] & \multicolumn{1}{l}{***} & [0.277015] & \multicolumn{1}{l}{**} \\
    $\beta_0$     & 11.99841 &       & 11.73815 &  \\
          & [3.252799] & \multicolumn{1}{l}{***} & [3.028546] & \multicolumn{1}{l}{**} \\
    $T$     & 0.076741 &       & 0.088838 &  \\
          & [0.007507] & \multicolumn{1}{l}{***} & [0.009463] & \multicolumn{1}{l}{**} \\
    $B_1$ & 0.564467 &       & 0.753142 &  \\
          & [0.056579] & \multicolumn{1}{l}{***} & [0.115291] & \multicolumn{1}{l}{**} \\
    $T(B_1)$ & -0.038034 &       & -0.05584 &  \\
          & [0.00612] & \multicolumn{1}{l}{***} & [0.01101] & \multicolumn{1}{l}{**} \\
    $B_2$ &       &       & -0.436834 &  \\
          &       &       & [0.239847] & \multicolumn{1}{l}{*} \\
    $T(B_2)$ &       &       & 0.014229 &  \\
          &       &       & [0.00755] & \multicolumn{1}{l}{*} \\
    \midrule
    \midrule
          &       &       &       &  \\
    \textbf{Adjusted $R^2$} & 0.99  &       & 0.99  &  \\
    \textbf{S.E. of Regression} & 0.04  &       & 0.04  &  \\
    \textbf{Long-run Variance} & 0.0014  &       & 0.0011  &  \\
    \textbf{Jarque-Bera Statistic} & 0.88  &       & 0.63  &  \\
    \bottomrule
    \end{tabular}
    \begin{tablenotes}
    \small
    \item Standard Deviations highlighted with * , ** and *** denote variables which are significant at 90\%, 95\% and 99\%, respectively. 
    \item Variables of first differences of regressors (as well as  their first lags) were not included in this results since they have only been used as a tool to estimate the long-run relationship between variables. Their coefficients do not have crucial interpretation. 
    \item The long-run variance was determined using a Bartlett kernel and fixed Newey-West bandwidth.
    \end{tablenotes}
  \label{Results6}
 \end{threeparttable}
\end{table}
\end{center}

It is important to mention, furthermore, that no relationship between EC and energy price was found. This results agrees with the finding of \protect\cite{j}, who did not find relationship between gasoline prices and gasoline  consumption by using data spanning the period 1979--2013, researching that fairly were carried out about an oil-based product. Unlike \protect\cite{j}, \protect\cite{i} found low gasoline demand price elasticities, however, such research was developed with information regarding the period 1989--1998. Noticeably, though until 1998 gasoline (which is an oil-based enegy source) was barely responsive to gasoline prices, towards 2013 such responsiveness had already decreasing at the point that no relationship existed anymore between consumption and prices of gasoline. The result of this research regarding price elasticity, therefore, can suggest that not just gasoline demand but also that of other energy sources suffered such decreasing in price responsiveness overtime.

\section{Conclusions and Policy Implications} \label{conclusions}

EC in Ecuador has kept an accelerated upward trend over the last years and, at the same time, has been modified significantly in its structure so that nowadays it is mainly comprised by oil-based EC (about the 80\% of the total) and it is mostly consumed by the transportation sector (about 50\% of the total). Given that Ecuador has become an oil-exporter country and an importer of oil-based products at the same time and, additionally, that large subsidies have been granted for the consumption of oil-based energy, during the last decades it has faced permanent problems of deficit budget. Thereby, EC issues are important topics to analyse by policymakers nowadays. The present research work sought to state the price and income elasticities of energy demand in Ecuador. 

The obtained results suggest, on one hand, that in the long-run the EC is significantly responsive to changes in real income as well as in industrial production level. As mentioned above, the trend shown by GDP level and EC specially during the last 15 years suggested the existence of a close relationship between the two series, such relationship was actually confirmed by the results of the fitted DOLS model. The responsiveness of EC to changes in industrial production, however, was found to be significant but low, which is a reasonable finding considering the participation of industrial sector on EC (around 18\% in 2014). Beyond that, such relationship was found to be negative, which can be explained by the fact that given the low level of capital investment in Ecuador, an increasing in industrial production might mean the mobilization of economic factors from other sectors to industry, so that EC by other sectors such as transportation would decrease, ending in a total negative effect on EC level. 

Regarding the price elasticity of energy demand, no significant relationship was found between energy price and EC. To this respect, about the price elasticity of gasoline demand, by one hand, \protect\cite{i} found low but significant gasoline demand price elasticities by using information spanning the period 1989--1998, while on the other hand, \protect\cite{j} did not find relationship between gasoline prices and gasoline consumption by using data spanning the period 1979--2013. Therefore, such previous results might suggest that while until 1998 gasoline was barely responsive to gasoline prices, towards 2013 such responsiveness had already decreasing at the point that no relationship existed anymore between consumption and gasoline prices. The result of this research regarding price elasticity of energy demand, therefore, suggests that not only gasoline demand but also that of other energy sources suffered such decreasing in price responsiveness overtime, ending up in a complete independence from the evolution of energy prices. 

Such a behavior regarding the relationship between EC and energy price, as \protect\cite{a} mentions, can be attributed to the substantial subsidies granted by the government in the acquirement of energy goods, given that subsidized energy price tends to distort energy demand level inducing excessive consumption and waste, and also discouraging the searching of more efficient and environmentally friendly energy sources. In Ecuador, in fact, consumption of several energy goods by various economic sectors, since 70's decade have been and are such largely subsidized that prices in internal market has remained relatively constant.

Important economic policy suggestions can be traced based on the results obtained in this research. As EC is highly sensitive to changes in real income, the establishment of income taxes as well as a better targeting of EC subsidies would (1) reduce the wasteful use of energy by the different economic sectors and (2) help  modifying the national budget dynamic, which during the last decades has presented large deficits due to the permanent trend of Ecuador---as oil-exporter country---to export of crude oil in exchange of importing oil-derivatives and the large subsidies provided to population for the consumption of energy goods. More specifically, a modification in the targeting and level of oil-based EC subsidies, might also work towards a convenient mobilization of factors from the transportation sector---whose development is actually strongly encouraged by large subsidies and which does not generate any added value---to the industry sector---which consumes oil-based energy in lower proportions and does generate added value.    

Finally, even though during the last years various efforts have been made by the government in order to advance in the replacement of oil-based energy sources by other more sustainable energy sources (mainly hidro-electricity), apparently prices of oil-based energy products in the internal market are as lower than their prices in the international market as they continue pressuring upward the internal demand of such goods, and so, the assignment of state funds to their imports. The participation of such goods remains around 80\% of the total EC. It is crucial for the country, however, and even more in the current international adverse context of oil price---wherein it bounds barely the \$45 so that value of crude oil exports of the Ecuador have decreased dramatically---that decisions to modify the current patterns of EC are taken in order to achieve a more financially sustainable performance of ecuadorian economy in the long run. In addition, rightly structured measures of economic policy taken to this respect could contribute towards a more ecologically sustainable development of the country.

\section*{Acknowledgement}

I thank Andrea Bonilla, Yasmín Salazar and Carolina Guevara, Professors at Escuela Politécnica Nacional, for their valuable suggestions and comments to improve this research.

\newpage
\nocite{bp,MCSE2014,MCSE2011,MinFinEc15,MinFinEc11}
\bibliographystyle{apalike}
\bibliography{biblio}

\end{document}